%% file: main.tex
\documentclass[sigconf,nonacm]{acmart}
\AtBeginDocument{%
  }

\setcopyright{none}

\usepackage{algorithm}
\usepackage{algpseudocode}

\usepackage{subcaption}

\begin{document}

\title{Coordinate Condensation: Subspace-Accelerated Coordinate Descent for Physics-Based Simulation}

\author{Ty Trusty}
\email{trusty@cs.toronto.edu}
\affiliation{%
  \institution{University of Toronto}
  \city{Toronto}
  \country{Canada}
}
\renewcommand{\shortauthors}{Trusty et al.}

\begin{abstract}
  We introduce \emph{Coordinate Condensation}, a variant of coordinate descent that accelerates physics-based simulation by augmenting local coordinate updates with a Schur-complement-based subspace correction.  
Recent work by Lan et al.~\cite{Lan2024} uses perturbation subspaces to augment local solves to account for global coupling, but their approach introduces damping that can degrade convergence.  
We reuse this subspace but solve for local and subspace displacements independently, eliminating this damping.  
For problems where the subspace adequately captures global coupling, our method achieves near-Newton convergence while retaining the efficiency and parallelism of coordinate descent.  
Through experiments across varying material stiffnesses and mesh resolutions, we show substantially faster convergence than both standard coordinate descent and JGS2~\cite{Lan2024}.  
We also characterize when subspace-based coordinate methods succeed or fail, offering insights for future solver design.

\end{abstract}



\begin{teaserfigure}
  \includegraphics[width=\textwidth]{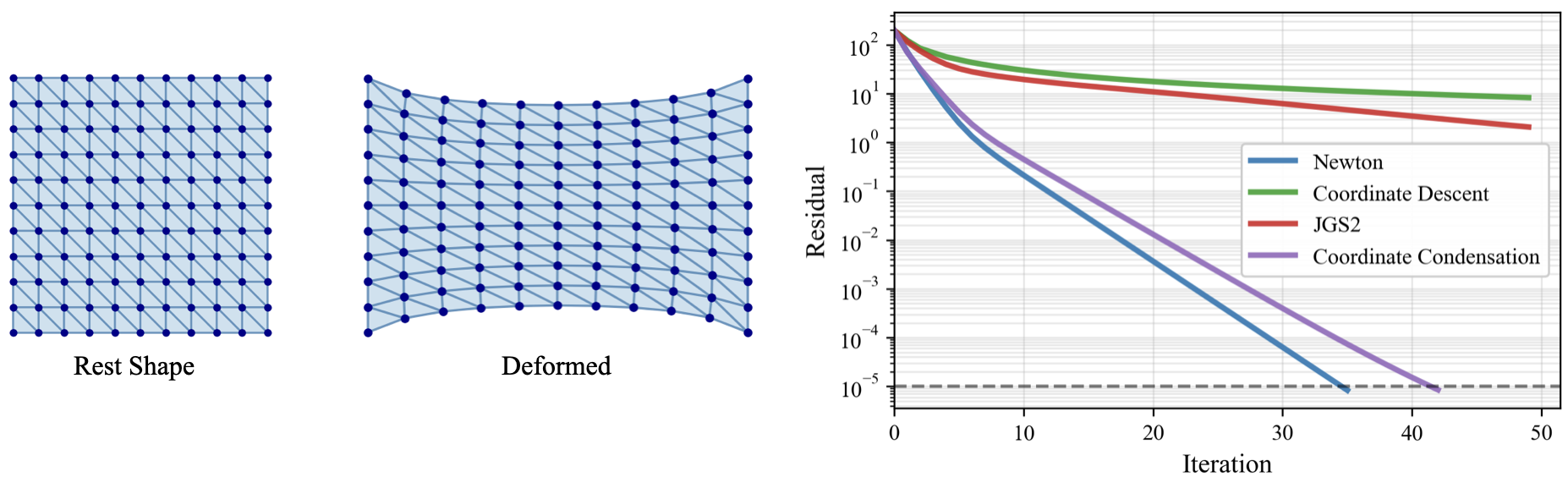}
  \caption{%
      Solver convergence comparison on a nonlinear quasistatic stretch of a square. Newton's method achieves 
      quadratic convergence, and while JGS2 \cite{Lan2024} and coordinate descent are slow to converge, our method, Coordinate Condensation, converges at a rate similar to Newton's method.
  }
  \label{fig:teaser}
\end{teaserfigure}


\maketitle

\input{intro}

\input{method}

\input{evaluations}
\input{conclusion}

\bibliographystyle{ACM-Reference-Format}
\bibliography{sample-base}

\end{document}

%% file: intro.tex
\section{Introduction}

Optimization-based solvers for implicit time integration of elastodynamics have become a standard tool in physics-based animation.  
Solving the resulting nonlinear optimization problems via Newton's method yields stable, high-fidelity simulations~\cite{Gast2015}, but the per-iteration cost (computing and inverting the full Hessian) can be prohibitive for large-scale or real-time applications.

To address this, recent work has explored \emph{Coordinate Descent} (CD)~\cite{Wright2015}, which decomposes the global problem into many small, parallelizable local subproblems, which enables efficient GPU implementations.
Notable recent examples include Vertex Block Descent (VBD)~\cite{Chen2024, Giles2025}, Second-Order Stencil Descent~\cite{Lan2023}, and JGS2~\cite{Lan2024}.
While these methods achieve impressive runtimes, their convergence can degrade when strong coupling due to stiff materials, fine mesh resolution, or constraints.
Local updates can only propagate information gradually, leading to slow convergence and often lagged dynamics. 

To address slow convergence caused by global coupling, Lan et al.~\cite{Lan2024} proposed JGS2, which augments each local coordinate update with a precomputed subspace representing the expected displacement of surrounding degrees of freedom (DOFs).  
Their method was designed to mitigate an ``overshooting'' effect observed in standard coordinate descent methods where local updates underestimate the global energy increase.
JGS2 enforces a rigid proportionality between the local update and the subspace displacement, effectively damping the local update.  
While this can improve stability in some cases, the fixed damping, as we show, can also degrade convergence.

\paragraph{Contributions.}
We introduce \emph{Coordinate Condensation}, which builds on JGS2's idea of using a per-vertex perturbation subspace to capture how a local degree of freedom influences the surrounding system.
Unlike JGS2, Coordinate Condensation decouples the local update and the subspace contribution, allowing the solver to balance local detail and global response.
This approach achieves consistently faster convergence across a range of problems with varying mesh resolution and material stiffness, while preserving the per-vertex structure and parallel efficiency of coordinate descent methods.

Our key contributions are:
\begin{itemize}
    \item \emph{Coordinate Condensation}, a coordinate descent solver with Schur-complement-based subspace correction.
    \item An evaluation of convergence across varying mesh resolutions, material stiffnesses, and subspace quality.
    \item A discussion of the basic limitations of subspace-based coordinate methods, offering directions for future solver design.
\end{itemize}

\section{Background}

\subsection{Optimization-Based Integration}

In physics-based animation, implicit time integration is often formulated in variational form where each time step advances the physics by solving an optimization problem~\cite{Gast2015}:
\begin{equation}
\mathbf{x}^{t+1} = \arg\min_{\mathbf{x}} E(\mathbf{x}),
\end{equation}
where $\mathbf{x} \in \mathbb{R}^{3n}$ represents the positions of $n$ vertices in 3D, and $E(\mathbf{x})$ is a nonlinear energy function typically composed of elastic, collision, and inertial potentials. For elastic simulation with implicit time integration, the energy takes the form:
\begin{equation}
E(\mathbf{x}) = \frac{1}{2}(\mathbf{x} - \tilde{\mathbf{x}})^T \mathbf{M} (\mathbf{x} - \tilde{\mathbf{x}}) + h^2 \Psi(\mathbf{x})
\end{equation}
where $\mathbf{M}$ is the mass matrix, $h$ is the time step, $\tilde{\mathbf{x}}$ \footnote{For Backward Euler time integration, $\tilde{\mathbf{x}} = \mathbf{x}^t + h \mathbf{v}^t$ with velocity $\mathbf{v}^t$.} is the inertial target, and $\Psi(\mathbf{x})$ is the elastic potential energy. Additional potentials can be added to the energy function to account for contact and other physical effects.

Newton's method solves this problem efficiently in terms of iterations, but requires inverting the full Hessian per step, which can be prohibitively expensive for large-scale problems. This motivates alternative solvers that trade some convergence rate for computational efficiency. While there is a plethora of approaches for this, we focus specifically on Coordinate Descent methods.

\subsection{Coordinate Descent}

Coordinate descent decomposes the global optimization into separate subproblems for individual coordinates (or small subsets thereof), allowing for efficient parallelization on GPUs where per-coordinate solves can be distributed over many threads. 

For a system with variables $\mathbf{x} = [x_1, x_2, \ldots, x_n]^T$, coordinate descent iteratively solves:
\begin{equation}
x_i^{k+1} = \arg\min_{x_i} E(x_1^k, \ldots, x_{i-1}^k, x_i, x_{i+1}^k, \ldots, x_n^k)
\end{equation}
for each coordinate $i$.
Variants of this include Jacobi-style in which updates are computed independently for each coordinate, and Gauss-Siedel-style where updates are computed using the most recent values for all coordinates. The former is more amenable to parallelization, while the latter generally yields better convergence.

\begin{figure}[t]
  \centering
  \includegraphics[width=0.48\textwidth]{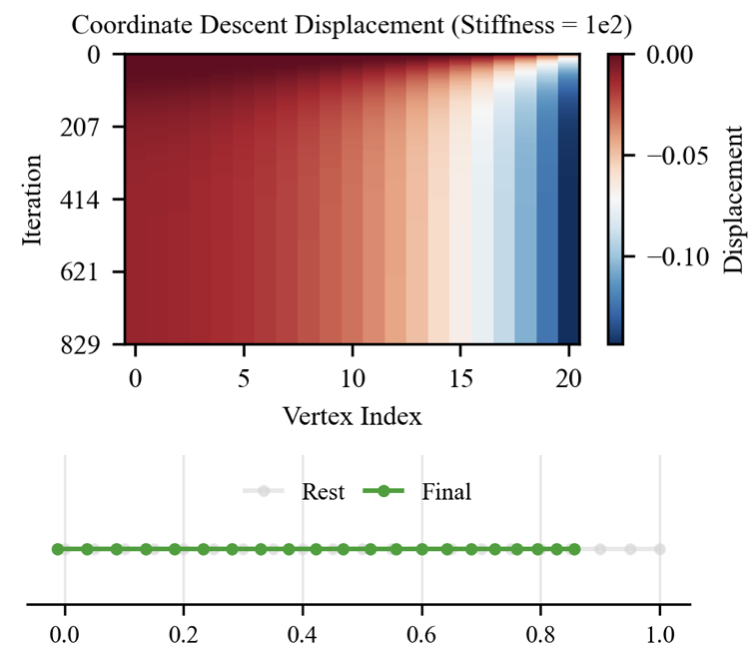}
  \caption{Slow convergence of coordinate descent on a 1D elastic rod with an applied impulse (right end). (Top) Heat map showing displacement magnitude across vertices (x-axis) and iterations (y-axis). Information propagates slowly from the impulse location, with vertices at the opposite end requiring many more iterations to reach their equilibrium displacement. (Bottom) Rest configuration (gray) and final configuration after one timestep.}
  \label{fig:cd_propagation}
\end{figure}

\begin{figure}[t]
  \centering
  \includegraphics[width=0.48\textwidth]{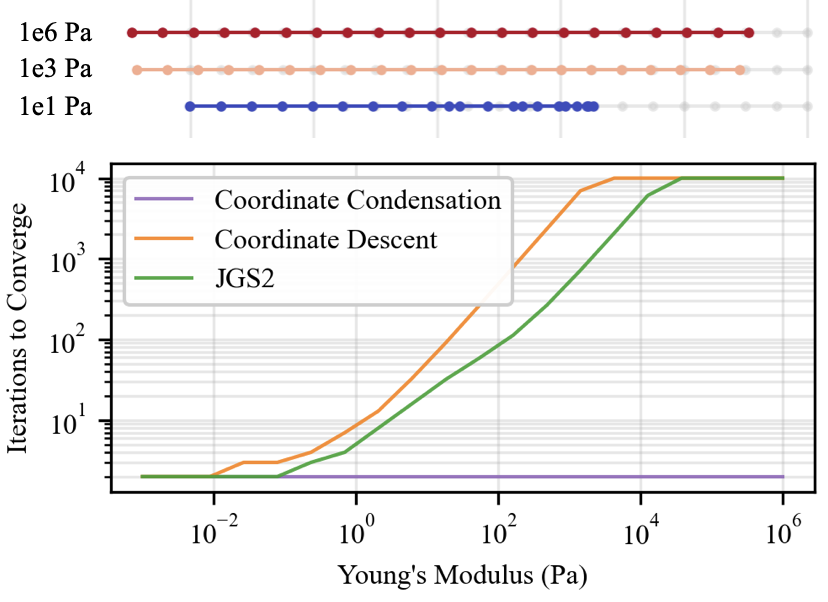}
  \caption{Convergence scaling with material stiffness for the 1D rod with impulse. (Top) Final configurations at three stiffness values showing the transition from mostly rigid translation (1e6 Pa) to localized deformation around the impulse location (1e1 Pa). (Bottom) Coordinate Condensation maintains constant iterations (optimal for this quadratic problem with the rest-shape basis), while Coordinate Descent and JGS2 both degrade significantly as stiffness increases, saturating the 10000 iteration limit. JGS2's damping provides minimal benefit over standard coordinate descent here.}
  \label{fig:rod_stiffness}
\end{figure}

\paragraph{Convergence characteristics.} 
While coordinate descent methods are efficient per iteration and highly parallelizable, their convergence rate depends heavily on the coupling strength between coordinates~\cite{Wright2015}.
For problems with weak coupling (e.g. off-diagonal Hessian elements are small relative to diagonals), convergence can be fast, but strong coupling leads to significant degradation.
In physics-based simulation, strong coupling is ubiquitous: mesh connectivity creates geometric coupling, while stiff materials amplify force propagation across the domain. Figure~\ref{fig:cd_propagation} illustrates this convergence behavior on a 1D elastic rod: vertices near the impulse converge rapidly, while distant vertices converge to their equilibrium displacement slowly.
As a result, convergence tends to break down as we increase problem size or material stiffness (Figure~\ref{fig:rod_stiffness}). The physical effect this has is that more iterations are required to propagate displacement across the domain, and this can manifest as lagged dynamics if iterations are limited per timestep (as is often the case in real-time solvers).

\subsection{JGS2: Accounting for Global Coupling}

\begin{figure}[t]
  \centering
  \includegraphics[width=0.48\textwidth]{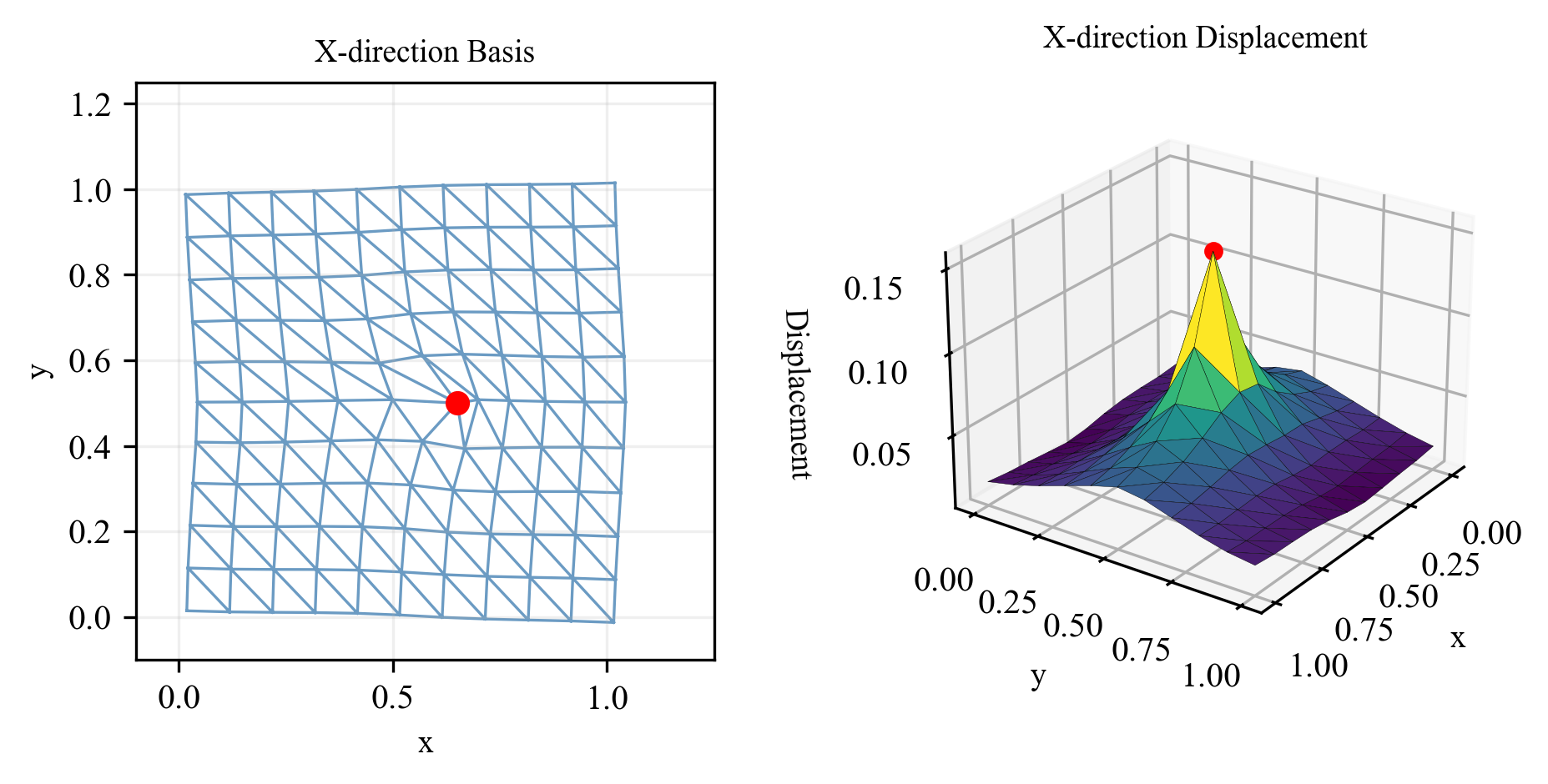}
  \caption{Perturbation basis $\mathbf{U}_{iC}$ for a single vertex (center, marked in red) in a 2D elastic mesh. (Left) 2D mesh with the perturbed vertex highlighted. (Right) Surface plot showing the x-direction displacement response of the mesh, where height and color indicate displacement magnitude.}
  \label{fig:basis}
\end{figure}

To address slow convergence due to strong global coupling, Lan et al.~\cite{Lan2024} proposed JGS2, which augments each coordinate update with a precomputed subspace capturing the expected response of complementary DOFs (DOFs other than the one being updated).
For each coordinate $i$, they build a perturbation basis $\mathbf{U}_{iC}$ representing how the complementary DOFs (all DOFs except $i$) would displace in response to a unit perturbation of coordinate $i$.
As a precomputation step, for each coordinate $i$, they assemble a block system:
\begin{equation}
\begin{bmatrix}
H_{ii} & H_{iC} \\
H_{Ci} & H_{CC}
\end{bmatrix}
\begin{bmatrix}
I \\
\mathbf{U}_{iC}
\end{bmatrix}
=
\begin{bmatrix}
\delta \mathbf{f}_i \\
\mathbf{0}
\end{bmatrix}
\end{equation}
where $I$ is the identity for coordinate $i$'s DOFs, $H_{CC}$ is the Hessian with respect to complementary DOFs, $H_{iC}$ couples coordinate $i$ to the complementary DOFs, and $\delta \mathbf{f}_i$ is a virtual force. This yields a subspace basis:
\begin{equation}
\mathbf{U}_{iC} = -H_{CC}^{-1} H_{Ci}
\end{equation}
representing how complementary DOFs respond to a perturbation of coordinate $i$ while maintaining equilibrium (zero force on complementary DOFs). Figure~\ref{fig:basis} visualizes this perturbation basis for a single vertex in a 2D mesh, showing how displacement propagates to neighboring vertices.

JGS2 formulates the local solve to include both the local coordinate and its influence on complementary DOFs. The full displacement is expressed for coordinate $i$ is:
\begin{equation}
\delta \mathbf{x} = \phi_i(\delta x_i) = \begin{bmatrix} I \\ \mathbf{U}_{iC} \end{bmatrix} \delta x_i
\end{equation}
which effectively displaces complementary DOFs proportionally to the local coordinate update ($\mathbf{U}_{iC} \delta x_i$ ). This then yields the local update:
\begin{equation}
\delta x_i = -\left(H_{ii} + \mathbf{U}_{iC}^T H_{CC} \mathbf{U}_{iC}\right)^{-1} \left(g_i + \mathbf{U}_{iC}^T \mathbf{g}_C\right)
\end{equation}
where $\mathbf{g}_C$ is the gradient with respect to complementary DOFs. This incorporates the stiffness and forces from the complementary subspace into the local solve, accounting for global coupling that standard coordinate descent ignores.

\paragraph{Limitations.} While JGS2 represents an improvement over standard coordinate descent in some scenarios, its formulation enforces a rigid proportionality between the coordinate update $\delta x_i$ and the complementary displacement $\mathbf{U}_{iC} \delta x_i$. This coupling is built into the structure of $\phi_i(\delta x_i)$---the complementary response is always exactly proportional to the local coordinate update, with $\mathbf{U}_{iC}$ determining the proportionality.

This rigid coupling effectively damps the local coordinate update. When the perturbation basis $\mathbf{U}_{iC}$ accurately captures the system's coupling structure, this can improve convergence. However, $\mathbf{U}_{iC} = -H_{CC}^{-1} H_{Ci}$ is computed at a reference configuration and may not accurately represent the coupling in the current deformed state. When the basis becomes inaccurate due to nonlinear deformations, large rotations, or unanticipated coupling (e.g., contact), the fixed damping often provides only meager improvements over standard coordinate descent, and in some cases can even degrade performance or fail to produce effective descent directions.

%% file: method.tex
\section{Method}

We now present the details of our Schur complement formulation, which enables the independent solving for local and complementary displacements. Algorithm 1 provides a high-level overview of the Jacobi variant.

\begin{algorithm}
\caption{Coordinate Condensation (Jacobi)}
\begin{algorithmic}[1]
\State \textbf{Precompute:} Perturbation basis $\mathbf{U}_i$ for each coordinate $i$ \Comment{See Section \ref{sec:coord-cond}}
\State $\mathbf{x} \gets \mathbf{x}^t$ 
\While{not converged}
    \State $\mathbf{g}, \mathbf{H} \gets \nabla E(\mathbf{x}), \nabla^2 E(\mathbf{x})$ 
    \For{$i = 1$ to $n$}
        \State Compute $S$ and $\tilde{g}_i$ \Comment{Eq.~\ref{eq:schur}}
        \State $\delta x_i, \delta \alpha_i \gets -(H_{ii} - S)^{-1} \tilde{g}_i$ \Comment{Eq.~\ref{eq:update}}
        \State $d_i \gets \delta x_i$
    \EndFor
    \State $\mathbf{x} \gets \mathbf{x} + \mathbf{d}$
\EndWhile
\State \Return $\mathbf{x}$
\end{algorithmic}
\end{algorithm}

Our approach is inspired by subspace condensation~\cite{Teng2015}. While Teng et al. augment subspace simulations with full-space detail, we augment per-coordinate solves with a complementary subspace that captures global coupling.
This maintains the parallelizability and per-iteration efficiency of coordinate descent while offering improved convergence rates over standard coordinate descent.

\subsection{Coordinate Condensation}\label{sec:coord-cond}

We precompute the perturbation basis $\mathbf{U}_i$ for each coordinate $i$ at the rest configuration by solving $\mathbf{U}_i = -H_{CC}^{-1} H_{Ci}$, following the same approach as JGS2.

Rather than enforcing a rigid proportionality between the local coordinate update and complementary displacement (as in JGS2), we solve for both quantities independently. We express the local displacement as:
\begin{equation}\label{eq:displacement}
\delta \mathbf{x}_i = \begin{bmatrix} I & 0 \\ 0 & \mathbf{U}_i \end{bmatrix} \begin{bmatrix} \delta x_i \\ \delta \alpha_i \end{bmatrix} = \mathbf{B}_i \mathbf{q}_i,
\end{equation}
where $\mathbf{U}_i$ is the perturbation basis for coordinate $i$, and $\mathbf{q}_i = [\delta x_i, \delta \alpha_i]^T$ contains the local coordinate and subspace displacements. With this displacement map, we can formulate the per-coordinate optimization as:
\begin{equation}\label{eq:optimization}
\min_{\mathbf{q}_i} E(\mathbf{x} + \delta \mathbf{x}_i) \approx E(\mathbf{x}) + \mathbf{g}^T \mathbf{B}_i \mathbf{q}_i + \frac{1}{2} \mathbf{q}_i^T \mathbf{B}_i^T \mathbf{H} \mathbf{B}_i \mathbf{q}_i,
\end{equation}
where $\mathbf{g}$ and $\mathbf{H}$ are the gradient and Hessian evaluated at $\mathbf{x}$.

Expanding the terms, we obtain the coupled block system,
\begin{equation}\label{eq:block}
\begin{bmatrix}
H_{ii} & H_{iC} \mathbf{U}_i \\
\mathbf{U}_i^T H_{iC}^T & \mathbf{U}_i^T H_{CC} \mathbf{U}_i
\end{bmatrix}
\begin{bmatrix}
\delta x_i \\
\delta \alpha_i
\end{bmatrix}
=
-\begin{bmatrix}
g_i \\
\mathbf{U}_i^T \mathbf{g}_C
\end{bmatrix}
\end{equation}
Through block elimination, this yields:
\begin{equation}\label{eq:update}
\delta x_i = -\left(H_{ii} - S\right)^{-1} \tilde{g}_i,
\end{equation}
where $H_{ii} - S$ is the Schur complement. Defining $\tilde{H}_{ii} = \mathbf{U}_i^T H_{CC} \mathbf{U}_i$ as the reduced complementary stiffness, we have:
\begin{align}\label{eq:schur}
S &= H_{iC} \mathbf{U}_i \tilde{H}_{ii}^{-1} \mathbf{U}_i^T H_{iC}^T \\
\tilde{g}_i &= g_i - H_{iC} \mathbf{U}_i \tilde{H}_{ii}^{-1} \mathbf{U}_i^T \mathbf{g}_C
\end{align}

\paragraph{Comparison to JGS2.} The structure of our update is similar to JGS2, but with a crucial difference: JGS2 uses $(H_{ii} + \mathbf{U}_i^T H_{CC} \mathbf{U}_i)$ for its update Hessian, which strictly adds stiffness to the system, always damping the update. In contrast, subtracting the Schur complement $S$ from $H_{ii}$ effectively deflates the stiffness by removing the component that couples to the complementary subspace, precisely accounting for the coupling rather than simply damping it. This allows the solver to take larger, more accurate steps when the subspace basis accurately captures the system's response. 
Our formulation has similar computational cost to JGS2, with the main added cost being the inversion of the small subspace matrix $\tilde{H}_{ii}$, which is negligible for small subspaces (a 2x2 solve in 2D, 3x3 in 3D).

\paragraph{Maintaining Locality.} Crucially, in our per-coordinate solve we only apply the update $\delta x_i$ to coordinate $i$, and discard the complementary subspace contribution $\mathbf{U}_i \delta \alpha_i$, and so the actual displacement of complementary DOFs is determined by their respective coordinate solves. This maintains the locality of updates, which is essential for practical implementation.

\paragraph{Accounting for Large Deformations.}
For nonlinear problems undergoing large deformations, the precomputed basis $\mathbf{U}_i$ can become stale as the configuration deviates from the rest state.
Following JGS2, we address this by estimating per-vertex rotations $\mathbf{R}_j \in SO(3)$ at each iteration using the displacement of the neighborhood around each vertex.
We then rotate each vertex's DOF block in the basis:
\begin{equation}\label{eq:rotation}
\mathbf{U}_i^{\text{rot}}[j] = \mathbf{R}_j \mathbf{U}_i[j]
\end{equation}
where $\mathbf{U}_i[j]$ denotes the DOF block corresponding to vertex $j$ in basis $\mathbf{U}_i$.
This co-rotated formulation maintains subspace quality without expensive basis reconstruction (see Figure~\ref{fig:subspace_quality}).

%% file: evaluations.tex
\section{Results}

\subsection{Implementation Details}

Our implementation is written in Python and uses the Simkit library~\cite{Simkit}. As we are concerned with convergence behavior rather than runtime performance, we do not implement the common optimizations (e.g. parallelization over each coordinate solve and cubature for full-space products \cite{Lan2024}).
While these optimizations are necessary for high-performance applications, our algorithm is equally amenable to such optimizations.
For the nonlinear elastic energy, we use the Fixed Corotational elasticity model~\cite{Stomakhin2012}.

\paragraph{Termination criteria.}
Unless otherwise mentioned, all experiments use a normalized gradient norm as the termination criterion: $\|\mathbf{g}\|/(V \cdot n \cdot E) < \epsilon$, where $V$ is total mesh volume, $n$ is the number of vertices, and $E$ is Young's modulus. This gives us a residual measure independent of how these quantities are scaled. For dynamic problems we also divide by the timestep size.

In the following experiments, we use the Jacobi variant of coordinate descent for all methods.

\subsection{Mesh Resolution Scaling}

\begin{figure}[t]
\centering
\includegraphics[width=0.48\textwidth]{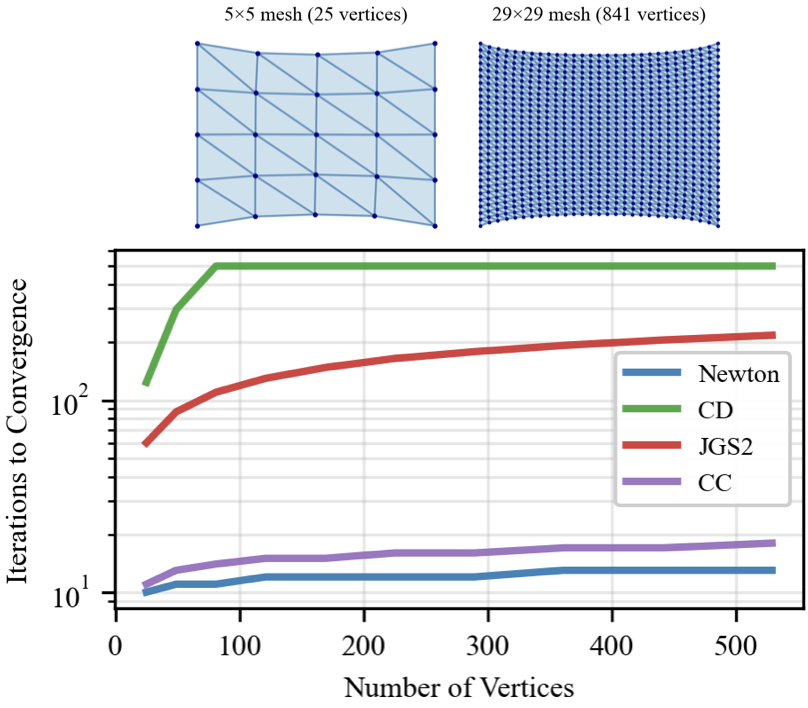}
\caption{Solver convergence as mesh resolution increases on the 2D elastic stretch test. (Top) Converged configurations for the coarsest mesh (5$\times$5, 25 vertices) and finest mesh (29$\times$29, 841 vertices) tested, showing the stretched deformation. (Bottom) Iterations to convergence as we increase the number of vertices. Standard Coordinate Descent quickly saturates the 500 iteration limit. JGS2 shows significant improvement but still scales poorly. Coordinate Condensation remains close to Newton's method, demonstrating near-optimal convergence across all resolutions.}
\label{fig:mesh_resolution}
\end{figure}

Figure~\ref{fig:mesh_resolution} compares Newton's method, standard Coordinate Descent, JGS2, and Coordinate Condensation as mesh resolution increases on the 2D elastic stretch test shown in Figure~\ref{fig:teaser}. The maximum iteration limit is set to 500 with a tolerance of $10^{-2}$.

Standard coordinate descent quickly saturates the maximum iteration count as the mesh is refined, highlighting the poor scaling characteristic of local methods. JGS2 shows a significant improvement by incorporating complementary stiffness, but still requires significantly more iterations than Newton's method. In contrast, Coordinate Condensation achieves convergence rates nearly matching Newton's method across all resolutions, demonstrating that the Schur complement formulation effectively captures the global coupling structure.

\subsection{Material Stiffness Scaling}

Figure~\ref{fig:rod_stiffness} examines convergence behavior as material stiffness varies, using the same 1D rod setup from Figure~\ref{fig:cd_propagation}.
For this experiment, we measure convergence using the criterion $\|\mathbf{x} - \mathbf{x}_{\text{Newton}}\|/n < 10^{-3}$, where $\mathbf{x}_{\text{Newton}}$ is the exact solution from Newton's method (solved to tight tolerance) and $n$ is the number of vertices. This is well-defined as the quadratic energy yields a unique solution.

The final configurations at different stiffness values reveal how material properties affect deformation characteristics.
At high stiffness (1e6 Pa), the rod largely undergoes rigid translation with minimal compression.
As stiffness decreases (1e3 Pa, 1e1 Pa), the deformation becomes increasingly localized around the impulse location on the right end.
Despite these varying deformation modes, as stiffness increases, the impulse-induced displacement propagates further along the rod, requiring more iterations for local methods to reach equilibrium.

Coordinate Condensation achieves optimal convergence (constant iterations) regardless of stiffness.
For this quadratic problem the Schur complement gives us the exact global response, resulting in single iteration convergence. In contrast, both standard Coordinate Descent and JGS2 exhibit severe degradation, eventually saturating the maximum iteration limit (10000) as stiffness approaches 1e5 Pa Young's modulus, which is still considerably soft for physics-based simulation.
Notably, JGS2's damping strategy offers little improvement over standard coordinate descent for this scenario.
\subsection{Subspace Quality}

While Coordinate Condensation achieves excellent convergence when the subspace basis accurately represents the coupling structure, the quality of this approximation can degrade under large deformations, particularly when elements undergo significant rotation.

\begin{figure*}[t]
\centering
\includegraphics[width=\textwidth]{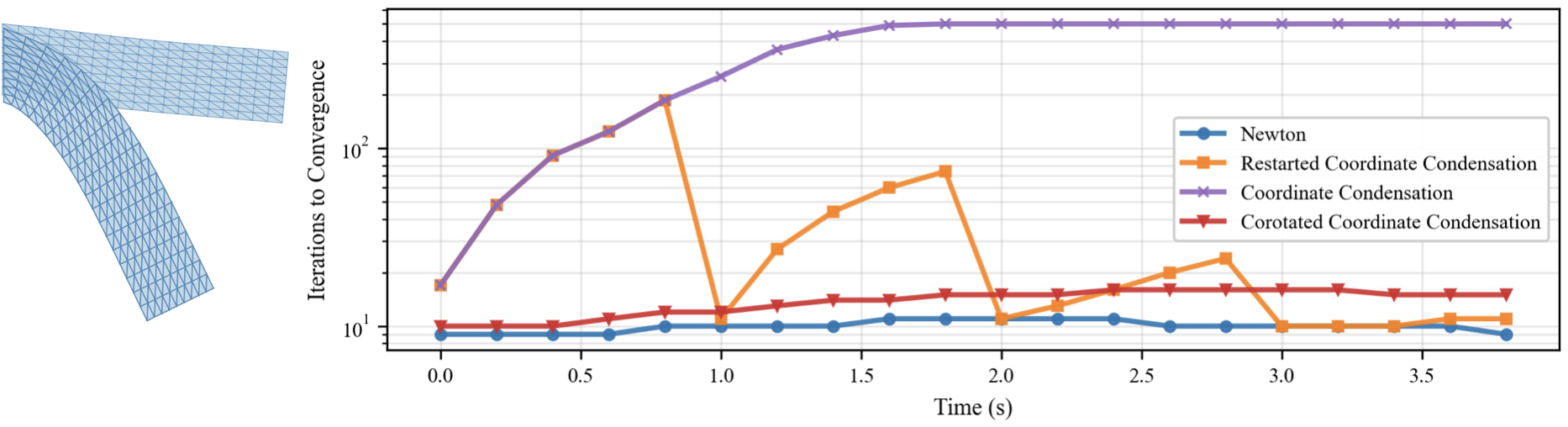}
\caption{Subspace quality analysis on a cantilever beam quasi-static simulation. (Left) Beam configurations throughout the simulation. (Right) Convergence comparison. Standard Coordinate Condensation with a fixed basis shows degraded convergence as the beam deforms and elements rotate. Restarted Coordinate Condensation (basis reconstructed every 5 steps) recovers the performance intermittently, confirming that convergence breakdown stems from subspace staleness. Corotated Coordinate Condensation, which rotates basis entries with estimated per-vertex rotations at each iteration, maintains good convergence without expensive basis reconstruction.}
\label{fig:subspace_quality}
\end{figure*}

\paragraph{Subspace degradation during large deformations.}
Figure~\ref{fig:subspace_quality} demonstrates this phenomenon using a cantilever beam undergoing quasi-static bending. We compare four variants: Newton's method, standard Coordinate Condensation with a fixed basis computed at initialization, Restarted Coordinate Condensation that reconstructs the subspace every 5 time steps, and Corotated Coordinate Condensation that rotates basis entries using estimated per-vertex rotations.

The results show that standard Coordinate Condensation with a fixed basis shows degraded convergence as the simulation progresses and the beam undergoes large deformation. The Restarted variant, which periodically reconstructs the subspace from the current configuration, sees improved convergence, but this approach would be prohibitively expensive in practice. Importantly it confirms that the convergence breakdown is a symptom of the subspace becoming stale.
More practically, the co-rotated variant maintains good convergence throughout the simulation without the computational cost of full basis reconstruction.

\begin{figure}[t]
  \centering
  \includegraphics[width=0.48\textwidth]{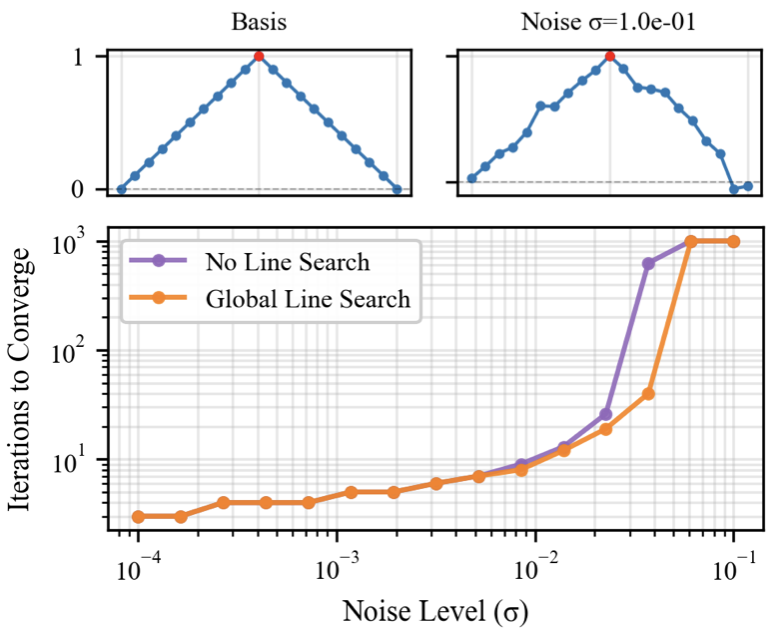}
  \caption{Sensitivity to basis quality on the 1D stretched rod. (Top) Visualization of the perturbation basis for the middle vertex: left shows the initial basis (compatible with boundary conditions), right shows basis perturbed with random noise $\sigma=0.1$. (Bottom) Iterations to converge as noise magnitude increases. Both standard Coordinate Condensation and the variant with global line search degrade significantly, eventually failing to converge within 1000 iterations at tolerance $10^{-2}$. While line search improves robustness, fundamental basis degradation limits convergence.}
  \label{fig:basis_noise}
\end{figure}

\paragraph{Sensitivity to basis quality.}
To further investigate the impact of subspace quality on convergence, we conduct an experiment using the 1D rod from Figures~\ref{fig:cd_propagation} and~\ref{fig:rod_stiffness}, but with prescribed boundary conditions at both ends and a stretch applied to the right end. Figure~\ref{fig:basis_noise} examines how artificially degrading the subspace basis affects convergence.

We perturb the basis by adding scaled random noise: $\mathbf{U}_{\text{noisy}} = \mathbf{U}_{\text{initial}} + \sigma \cdot \mathbf{1}$, where $\sigma$ controls the noise magnitude. We test two variants: Coordinate Condensation with and without a global line search. The line search variant accumulates the descent directions from all per-coordinate solves into a global descent direction, $\mathbf{d}$, then performs backtracking line search to find the optimal step size $\alpha^* = \arg\min_\alpha E(\mathbf{x} + \alpha \mathbf{d})$ before updating. This ideally ensures energy reduction despite basis inaccuracies.

The results show as noise increases, both variants degrade significantly. While the line search improves robustness and reduces iterations for moderate noise levels, both methods eventually fail to converge within 1000 iterations as the basis quality deteriorates. So while line search can partially compensate for basis errors, degradation of the subspace still limits convergence.

\subsection{Nonlinear Problems}

\begin{figure}[t]
  \centering
  \includegraphics[width=0.48\textwidth]{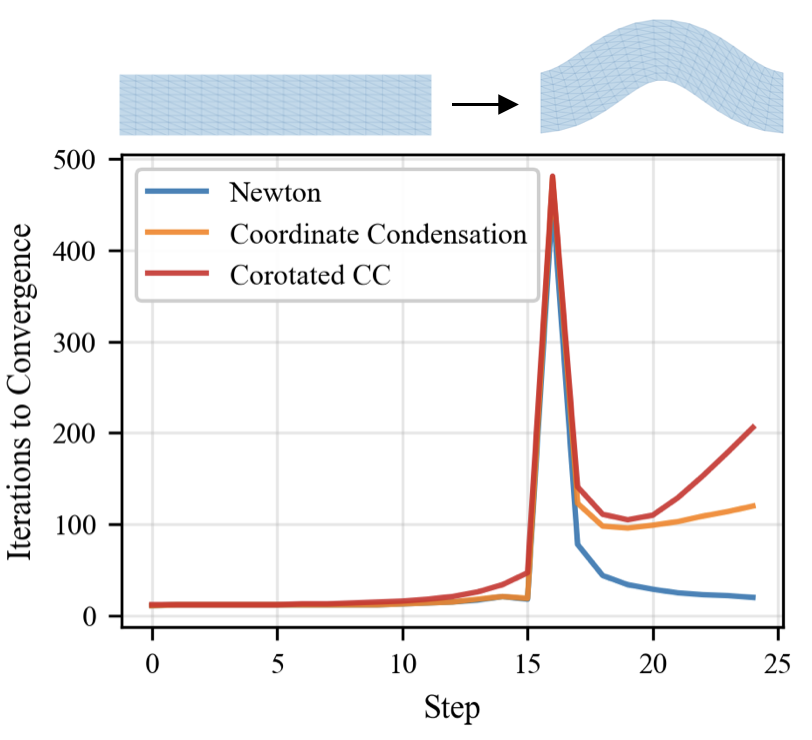}
  \caption{Buckling simulation demonstrating challenging nonlinear behavior. (Top) Initial and final configurations of the beam. (Bottom) Iterations to converge at each quasistatic step. The buckling point (around step 15) is challenging for all methods. Before buckling, both Coordinate Condensation and Corotated CC track Newton closely. After buckling, both methods deviate from Newton, with Corotated CC suffering more, suggesting that the rotation-based adaptation does not capture the missing nonlinearity well in this scenario.}
  \label{fig:buckling}
\end{figure}

To examine performance under extreme nonlinearity, we perform a quasistatic buckling simulation where boundary conditions are progressively updated to compress a beam until it buckles (Figure~\ref{fig:buckling}).

Up until the buckling point, both Coordinate Condensation and Corotated Coordinate Condensation follow Newton's method closely in terms of iterations required.
The buckling step itself requires many iterations for all methods, including Newton, as this represents a highly nonlinear transition.
After buckling, both coordinate descent variants deviate from Newton's convergence rate.
Notably, Corotated CC suffers more than standard Coordinate Condensation after the buckling point, suggesting that the per-vertex rotation estimates do not adequately capture the highly nonlinear response in the buckled configuration.

\begin{figure*}[t]
  \centering
  \includegraphics[width=\textwidth]{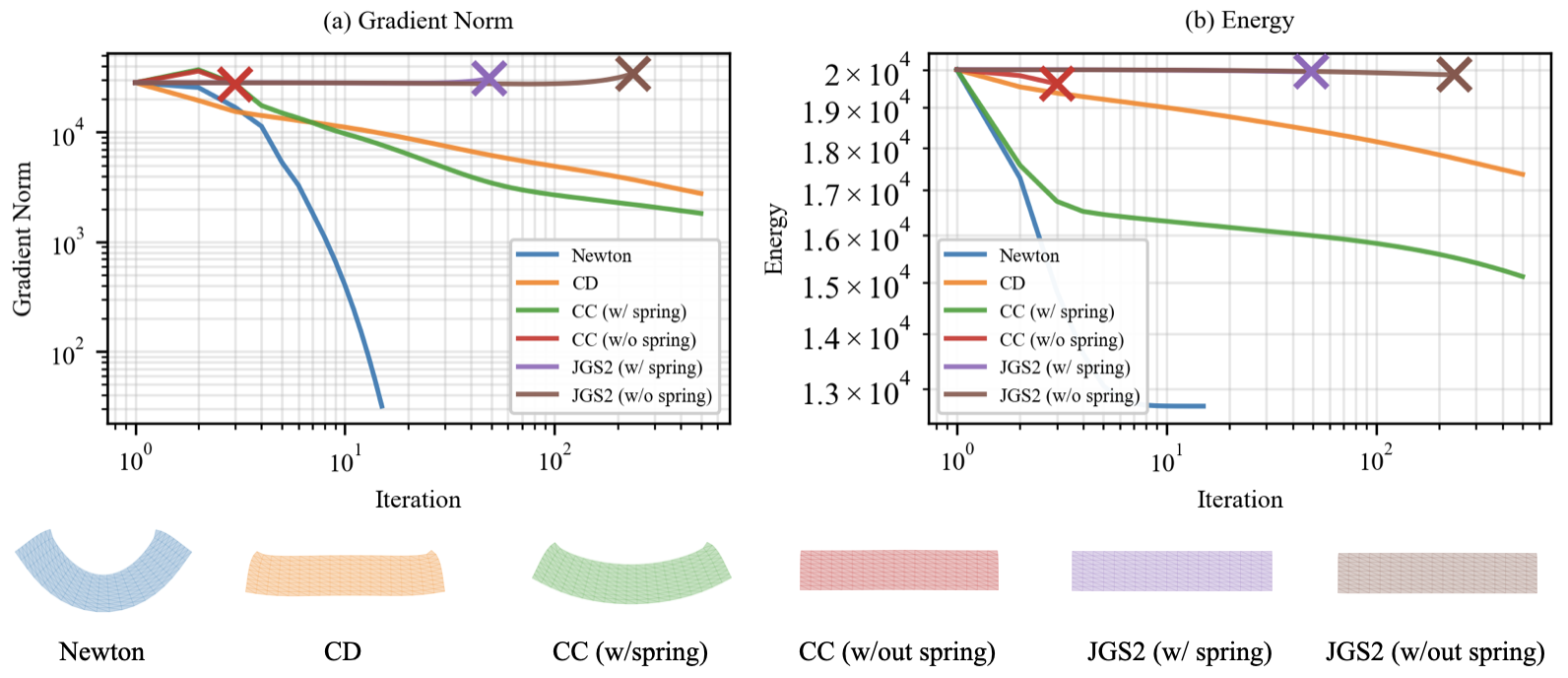}
  \caption{Convergence with unanticipated coupling. A spring connects the top corners of the beam, inducing bending. We compare standard Coordinate Descent (CD), Coordinate Condensation (CC), and JGS2, where the spring energy is either included (w/ spring) or excluded (w/o spring) during basis construction for CC and JGS2. (a) Gradient norm shows Newton converging fastest. Among coordinate descent methods, CC w/ spring converges to a lower gradient norm faster than CD, while JGS2 w/ spring completely stagnates, unable to produce a descent direction. Both CC w/o spring and JGS2 w/o spring fail to converge, demonstrating that when the precomputed basis lacks awareness of coupling, the subspace methods cannot recover. (b) Energy plots confirm CC w/ spring reduces energy effectively while other subspace variants fail. Final configurations (bottom) show only CC w/ spring reaches a result resembling Newton's solution.}
  \label{fig:spring}
\end{figure*}

\paragraph{Unanticipated coupling.}
Figure~\ref{fig:spring} demonstrates a critical limitation of precomputed subspace methods: when new coupling emerges during simulation that was not captured in the precomputed basis.
We attach a spring between the top left and top right vertices of a beam, inducing bending, and perform a single quasistatic solve with all solvers capped at 500 iterations.
To prevent divergence in this scenario, we all coordinate descent solver use a global line search at each iteration.
This experiment serves as a proxy for effects like contact which introduce new coupling terms in the Hessian during simulation.

The results reveal the critical dependence on basis quality.
When the spring coupling is included during basis construction (CC w/ spring), Coordinate Condensation converges to a lower energy much faster than other Coordinate Descent methods.
However, JGS2 w/ spring completely stagnates—unable, even with the spring accounted for in its basis.
When the spring energy is excluded from basis construction (CC w/o spring and JGS2 w/o spring), both subspace methods fail to converge entirely.
The final configurations confirm that only CC w/ spring reaches a solution resembling Newton's result.

This experiment highlights a fundamental limitation of precomputed subspace methods: they are inherently limited by the coupling structure present during basis construction.
When new interactions emerge during simulation (whether from contact, springs, or other dynamic constraints) methods that rely rigidly on precomputed bases can fail catastrophically.
While our Coordinate Condensation approach shows promising results when the basis captures relevant coupling, this highlights the need for future work on adaptive strategies.


%% file: conclusion.tex
\section{Conclusion and Future Work}

We have presented Coordinate Condensation, a coordinate descent method that incorporates a Schur-complement-based subspace correction to local coordinate solves.
Using the same per-vertex perturbation basis as JGS2, our method solves for both local vertex updates and complementary subspace displacements independently, removing the rigid proportionality that damps JGS2's convergence.
Our experiments demonstrate substantially improved convergence compared to both standard coordinate descent and JGS2 across problems with varying mesh resolution and material stiffness, including optimal (constant iteration) convergence for quadratic problems.

However, like all subspace methods, Coordinate Condensation is fundamentally limited by the quality and relevance of the precomputed basis.
When new coupling emerges during simulation (e.g., contact, Figure~\ref{fig:spring}), when bases become stale under large deformations (Figure~\ref{fig:subspace_quality}), or when extreme nonlinearity arises (Figure~\ref{fig:buckling}), convergence degrades.
While our Schur complement formulation provides advantages over JGS2's fixed proportionality in these scenarios, both methods ultimately depend on the basis accurately capturing the system's coupling.

Future work could address these limitations through adaptive strategies: detecting when new coupling emerges and either updating bases or augmenting them with additional modes; error estimation to trigger basis updates~\cite{Trusty2024} or fallback to standard coordinate descent.


%% file: main.bbl

\begin{thebibliography}{10}


\ifx \showCODEN    \undefined \def \showCODEN     #1{\unskip}     \fi
\ifx \showISBNx    \undefined \def \showISBNx     #1{\unskip}     \fi
\ifx \showISBNxiii \undefined \def \showISBNxiii  #1{\unskip}     \fi
\ifx \showISSN     \undefined \def \showISSN      #1{\unskip}     \fi
\ifx \showLCCN     \undefined \def \showLCCN      #1{\unskip}     \fi
\ifx \shownote     \undefined \def \shownote      #1{#1}          \fi
\ifx \showarticletitle \undefined \def \showarticletitle #1{#1}   \fi
\ifx \showURL      \undefined \def \showURL       {\relax}        \fi
\providecommand\bibfield[2]{#2}
\providecommand\bibinfo[2]{#2}
\providecommand\natexlab[1]{#1}
\providecommand\showeprint[2][]{arXiv:#2}

\bibitem[Benchekroun(2024)]%
        {Simkit}
\bibfield{author}{\bibinfo{person}{Otman Benchekroun}.}
  \bibinfo{year}{2024}\natexlab{}.
\newblock \bibinfo{title}{SimKit: A Simulation Toolkit for Computer Animation}.
\newblock \bibinfo{howpublished}{\url{https://github.com/otmanon/simkit}}.
\newblock
\newblock
\shownote{Accessed: 2025}.


\bibitem[Chen et~al\mbox{.}(2024)]%
        {Chen2024}
\bibfield{author}{\bibinfo{person}{Anka~He Chen}, \bibinfo{person}{Ziheng Liu},
  \bibinfo{person}{Yin Yang}, {and} \bibinfo{person}{Cem Yuksel}.}
  \bibinfo{year}{2024}\natexlab{}.
\newblock \showarticletitle{Vertex Block Descent}.
\newblock \bibinfo{journal}{\emph{ACM Trans. Graph.}} \bibinfo{volume}{43},
  \bibinfo{number}{4}, Article \bibinfo{articleno}{116} (\bibinfo{date}{July}
  \bibinfo{year}{2024}), \bibinfo{numpages}{16}~pages.
\newblock
\showISSN{0730-0301}
\href{https://doi.org/10.1145/3658179}{doi:\nolinkurl{10.1145/3658179}}


\bibitem[Gast and Schroeder(2015)]%
        {Gast2015}
\bibfield{author}{\bibinfo{person}{T.~F. Gast} {and} \bibinfo{person}{C.
  Schroeder}.} \bibinfo{year}{2015}\natexlab{}.
\newblock \showarticletitle{Optimization integrator for large time steps}. In
  \bibinfo{booktitle}{\emph{Proceedings of the ACM SIGGRAPH/Eurographics
  Symposium on Computer Animation}} (Copenhagen, Denmark)
  \emph{(\bibinfo{series}{SCA '14})}. \bibinfo{publisher}{Eurographics
  Association}, \bibinfo{address}{Goslar, DEU}, \bibinfo{pages}{31--40}.
\newblock


\bibitem[Giles et~al\mbox{.}(2025)]%
        {Giles2025}
\bibfield{author}{\bibinfo{person}{Chris Giles}, \bibinfo{person}{Elie Diaz},
  {and} \bibinfo{person}{Cem Yuksel}.} \bibinfo{year}{2025}\natexlab{}.
\newblock \showarticletitle{Crazy Fast Physics! Augmented Vertex Block Descent
  in Action!}. In \bibinfo{booktitle}{\emph{Proceedings of the Special Interest
  Group on Computer Graphics and Interactive Techniques Conference Real-Time
  Live!}} \emph{(\bibinfo{series}{SIGGRAPH Real-Time Live! '25})}.
  \bibinfo{publisher}{Association for Computing Machinery},
  \bibinfo{address}{New York, NY, USA}, Article \bibinfo{articleno}{3},
  \bibinfo{numpages}{2}~pages.
\newblock
\showISBNx{9798400715457}
\href{https://doi.org/10.1145/3721243.3735982}{doi:\nolinkurl{10.1145/3721243.3735982}}


\bibitem[Lan et~al\mbox{.}(2023)]%
        {Lan2023}
\bibfield{author}{\bibinfo{person}{Lei Lan}, \bibinfo{person}{Minchen Li},
  \bibinfo{person}{Chenfanfu Jiang}, \bibinfo{person}{Huamin Wang}, {and}
  \bibinfo{person}{Yin Yang}.} \bibinfo{year}{2023}\natexlab{}.
\newblock \showarticletitle{Second-order Stencil Descent for Interior-point
  Hyperelasticity}.
\newblock \bibinfo{journal}{\emph{ACM Trans. Graph.}} \bibinfo{volume}{42},
  \bibinfo{number}{4}, Article \bibinfo{articleno}{108} (\bibinfo{date}{July}
  \bibinfo{year}{2023}), \bibinfo{numpages}{16}~pages.
\newblock
\showISSN{0730-0301}
\href{https://doi.org/10.1145/3592104}{doi:\nolinkurl{10.1145/3592104}}


\bibitem[Lan et~al\mbox{.}(2025)]%
        {Lan2024}
\bibfield{author}{\bibinfo{person}{Lei Lan}, \bibinfo{person}{Zixuan Lu},
  \bibinfo{person}{Chun Yuan}, \bibinfo{person}{Weiwei Xu},
  \bibinfo{person}{Hao Su}, \bibinfo{person}{Huamin Wang},
  \bibinfo{person}{Chenfanfu Jiang}, {and} \bibinfo{person}{Yin Yang}.}
  \bibinfo{year}{2025}\natexlab{}.
\newblock \showarticletitle{JGS2: Near Second-order Converging
  Jacobi/Gauss-Seidel for GPU Elastodynamics}.
\newblock \bibinfo{journal}{\emph{ACM Trans. Graph.}} \bibinfo{volume}{44},
  \bibinfo{number}{4}, Article \bibinfo{articleno}{44} (\bibinfo{date}{July}
  \bibinfo{year}{2025}), \bibinfo{numpages}{15}~pages.
\newblock
\showISSN{0730-0301}
\href{https://doi.org/10.1145/3731183}{doi:\nolinkurl{10.1145/3731183}}


\bibitem[Stomakhin et~al\mbox{.}(2012)]%
        {Stomakhin2012}
\bibfield{author}{\bibinfo{person}{Alexey Stomakhin}, \bibinfo{person}{Russell
  Howes}, \bibinfo{person}{Craig Schroeder}, {and} \bibinfo{person}{Joseph~M.
  Teran}.} \bibinfo{year}{2012}\natexlab{}.
\newblock \showarticletitle{Energetically consistent invertible elasticity}. In
  \bibinfo{booktitle}{\emph{Proceedings of the ACM SIGGRAPH/Eurographics
  Symposium on Computer Animation}} (Lausanne, Switzerland)
  \emph{(\bibinfo{series}{SCA '12})}. \bibinfo{publisher}{Eurographics
  Association}, \bibinfo{address}{Goslar, DEU}, \bibinfo{pages}{25--32}.
\newblock
\showISBNx{9783905674378}


\bibitem[Teng et~al\mbox{.}(2015)]%
        {Teng2015}
\bibfield{author}{\bibinfo{person}{Yun Teng}, \bibinfo{person}{Mark Meyer},
  \bibinfo{person}{Tony DeRose}, {and} \bibinfo{person}{Theodore Kim}.}
  \bibinfo{year}{2015}\natexlab{}.
\newblock \showarticletitle{Subspace condensation: full space adaptivity for
  subspace deformations}.
\newblock \bibinfo{journal}{\emph{ACM Trans. Graph.}} \bibinfo{volume}{34},
  \bibinfo{number}{4}, Article \bibinfo{articleno}{76} (\bibinfo{date}{July}
  \bibinfo{year}{2015}), \bibinfo{numpages}{9}~pages.
\newblock
\showISSN{0730-0301}
\href{https://doi.org/10.1145/2766904}{doi:\nolinkurl{10.1145/2766904}}


\bibitem[Trusty et~al\mbox{.}(2024)]%
        {Trusty2024}
\bibfield{author}{\bibinfo{person}{Ty Trusty}, \bibinfo{person}{Yun~(Raymond)
  Fei}, \bibinfo{person}{David Levin}, {and} \bibinfo{person}{Danny Kaufman}.}
  \bibinfo{year}{2024}\natexlab{}.
\newblock \showarticletitle{Trading Spaces: Adaptive Subspace Time Integration
  for Contacting Elastodynamics}.
\newblock \bibinfo{journal}{\emph{ACM Trans. Graph.}} \bibinfo{volume}{43},
  \bibinfo{number}{6}, Article \bibinfo{articleno}{227} (\bibinfo{date}{nov}
  \bibinfo{year}{2024}), \bibinfo{numpages}{16}~pages.
\newblock
\showISSN{0730-0301}
\href{https://doi.org/10.1145/3687946}{doi:\nolinkurl{10.1145/3687946}}


\bibitem[Wright(2015)]%
        {Wright2015}
\bibfield{author}{\bibinfo{person}{Stephen~J. Wright}.}
  \bibinfo{year}{2015}\natexlab{}.
\newblock \bibinfo{title}{Coordinate Descent Algorithms}.
\newblock
\showeprint[arxiv]{1502.04759}~[math.OC]
\urldef\tempurl%
\url{https://arxiv.org/abs/1502.04759}
\showURL{%
\tempurl}


\end{thebibliography}
